\documentclass{article}

\usepackage{microtype}
\usepackage{graphicx}
\usepackage{subfigure}
\usepackage{booktabs} % for professional tables
\usepackage{multirow}
\usepackage{hyperref}
\usepackage{pdfpages}

\usepackage[accepted]{icml2023}

\usepackage{amsmath}
\usepackage{amssymb}
\usepackage{mathtools}
\usepackage{amsthm}

\usepackage[capitalize,noabbrev]{cleveref}

\theoremstyle{plain}

\theoremstyle{definition}

\theoremstyle{remark}

\usepackage[textsize=tiny]{todonotes}

\icmltitlerunning{Molecule-Morphology Contrastive Pretraining for Transferable Molecular Representation}

\begin{document}

\twocolumn[
\icmltitle{Molecule-Morphology Contrastive Pretraining for\linebreak Transferable Molecular Representation}




\begin{icmlauthorlist}
\icmlauthor{Cuong Q. Nguyen}{gskai}
\icmlauthor{Dante Pertusi}{gskmd}
\icmlauthor{Kim M. Branson}{gskai}
\end{icmlauthorlist}

\icmlaffiliation{gskai}{GSK, Artificial Intelligence and Machine Learning}
\icmlaffiliation{gskmd}{GSK, Medicine Design}

\icmlcorrespondingauthor{Cuong Q. Nguyen}{cuong.q.nguyen@gsk.com}

\icmlkeywords{Machine Learning, ICML}

\vskip 0.3in
]



\printAffiliationsAndNotice{}  % leave blank if no need to mention equal contribution

\begin{abstract}
Image-based profiling techniques have become increasingly popular over the past decade for their applications in target identification, mechanism-of-action inference, and assay development. These techniques have generated large datasets of cellular morphologies, which are typically used to investigate the effects of small molecule perturbagens. In this work, we extend the impact of such dataset to improving quantitative structure-activity relationship (QSAR) models by introducing Molecule-Morphology Contrastive Pretraining (MoCoP), a framework for learning multi-modal representation of molecular graphs and cellular morphologies. We scale MoCoP to approximately 100K molecules and 600K morphological profiles using data from the JUMP-CP Consortium and show that MoCoP consistently improves performances of graph neural networks (GNNs) on molecular property prediction tasks in ChEMBL20 across all dataset sizes. The pretrained GNNs are also evaluated on internal GSK pharmacokinetic data and show an average improvement of 2.6\% and 6.3\% in AUPRC for full and low data regimes, respectively. Our findings suggest that integrating cellular morphologies with molecular graphs using MoCoP can significantly improve the performance of QSAR models, ultimately expanding the deep learning toolbox available for QSAR applications. 

\end{abstract}
\begin{figure*}
\begin{center}
\includegraphics[width=\textwidth]{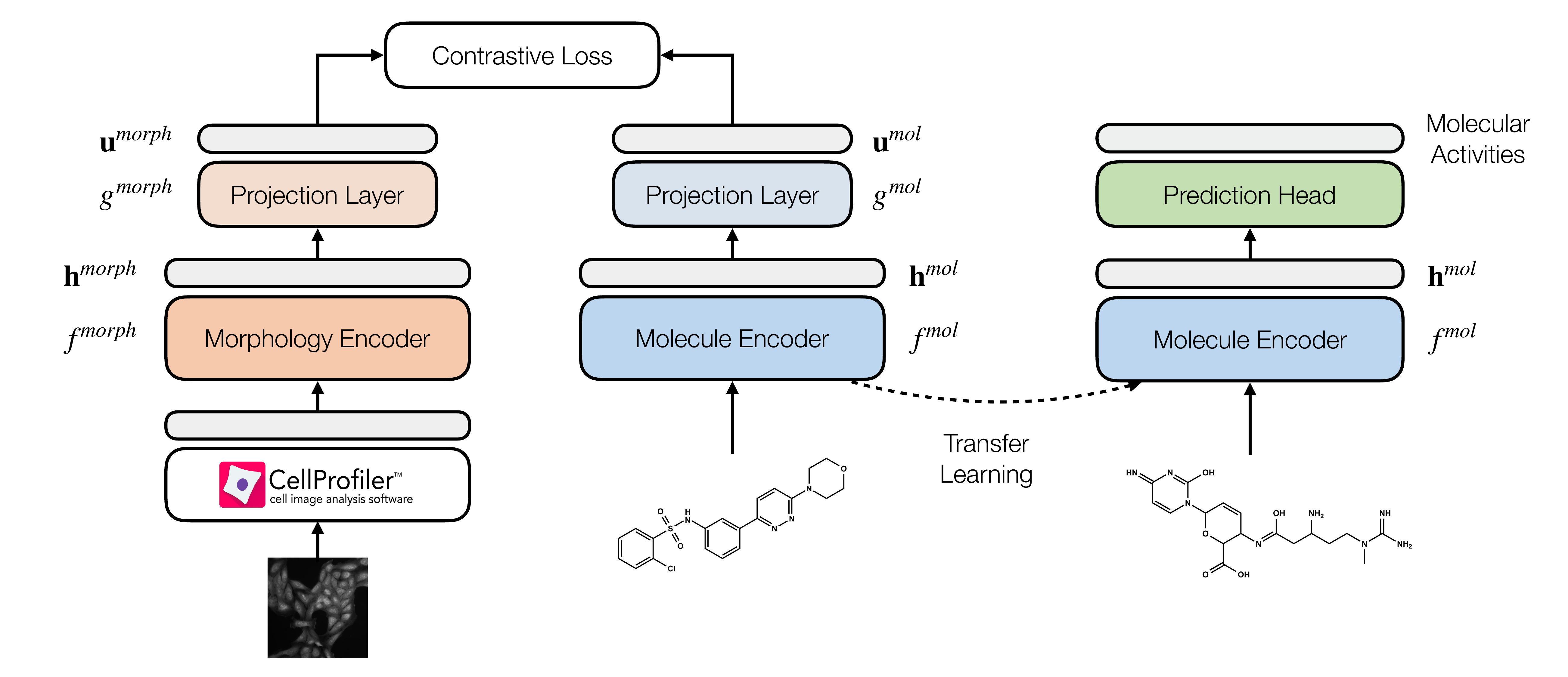}
\vskip -0.1in
\caption{Molecule-morphology contrastive learning workflow. We first jointly learn a molecule encoder and morphology encoder using contrastive learning on paired (molecule, morphology) data in available in the JUMP-CP dataset (left). Transfer learning is then performed by fine-tuning the pretrained molecule encoder on specific downstream tasks (right).}
\label{figure:retrieval}
\end{center}
\vskip -0.1in
\end{figure*}
\section{Introduction}
\label{introduction}
Quantitative structure-activity relationship (QSAR) modeling is a critical step for virtual screening in drug discovery, helping researchers prioritize modifications to chemical structures that shift modeled properties in a favorable direction. Since the Merck Molecular Activity Challenge, applying deep learning techniques to QSAR modeling has gained significant attention due to their ability to extract complex nonlinear relationships between chemical structures and their associated activities. Typically, QSAR models are trained to predict the activity of a molecule based on its in silico representation, which can have varying levels of complexity ranging from computed chemical properties, 2- and 3-D descriptors \cite{rogers_extended-connectivity_2010,sheridan_chemical_1996,carhart_atom_1985,nilakantan_topological_1987,schaller_next_2020}, and molecular graphs \cite{kearnes_molecular_2016,yang_analyzing_2019}.

However, performance of QSAR models is limited by the amount of available data, especially when assays are low-throughput, expensive to run, or only commissioned at the later stages of the drug discovery process. To overcome this limitation, methods such as active learning \cite{reker_active-learning_2015,smith_less_2018}, large-scale multitask learning \cite{xu_demystifying_2017,ramsundar_massively_2015,kearnes_modeling_2017} pretraining \cite{hu_strategies_2020}, and few-shot learning approaches \cite{altae-tran_low_2017,nguyen_meta-learning_2020} have been shown to improve model performance in low data regime.

Improving the in silico representation of molecules can also enhance performance of QSAR models. Recent trends in small-molecule drug discovery have shifted toward high-content screening approaches, with cellular imaging emerging as a relatively high-throughput \cite{kurita_connecting_2015,kraus_automated_2017,chandrasekaran_image-based_2021} method to profiling small molecules in relevant biological system. The Cell Painting assay \cite{bray_cell_2016} -- an unbiased and scalable approach for capturing images of cells -- have made large and reusable repositories of paired molecule and cell images possible \cite{bray_dataset_2017,fay_rxrx3_2023,chandrasekaran_jump_2023}. These images contain cellular morphologies induced by small molecule perturbagens and can be used as an alternative in silico representation of these molecules \cite{kraus_automated_2017,godinez_unsupervised_2018,hofmarcher_accurate_2019,stirling_cellprofiler_2021}. Convolutional neural network-based approaches have been shown to improve the predictivity of QSAR models across a wide range of assays \cite{hofmarcher_accurate_2019}, leading to increased hit rates and optimization of compounds to elicit a desired phenotype \cite{cuccarese_functional_2020}. However, the use of such models is limited by two factors: (1) cellular images are commonly plagued by batch effects, requiring extensive engineering efforts to learn domain agnostic representation \cite{ando_improving_2017,sypetkowski_rxrx1_2023}, and (2) only molecules that have paired cellular images can be used as input during inference, restricting the application of these models in virtual screening scenarios where such images are not available for the majority of molecules.

In parallel, contrastive learning has been shown to be effective for learning representations of multi-modal data. ConVIRT \cite{zhang_contrastive_2020} uses a modified InfoNCE objective \cite{oord_representation_2019} to learn a joint embedding space of medical images and human annotations. CLIP \cite{radford_learning_2021} scales up this approach to 400M (image, text) pairs, enabling zero-shot transfer to downstream image classification tasks. Recently, CLOOME \cite{sanchez-fernandez_contrastive_2022} uses the InfoLOOB objective \cite{furst_cloob_2022} to jointly learn a molecule encoder and a morphology encoder for molecular retrieval task using the dataset introduced by Bray et al. \yrcite{bray_dataset_2017}. Using the same dataset, Zheng et al. \yrcite{zheng_cross-modal_2022} extends this approach by including masked-graph modeling objective for pretraining graph neural networks (GNNs), showing improved performances on downstream tasks in the Open Graph Benchmark \cite{hu_open_2021}.

In this work, we further demonstrate the scaling of GNN-based \textbf{M}olecule-m\textbf{o}rphology \textbf{Co}ntrastive \textbf{P}retraining -- refered to as \textbf{MoCoP} -- from 30K molecules and 120K images in Bray et al. \yrcite{bray_dataset_2017} to approximately 100K molecules and 600K images in JUMP-CP \cite{chandrasekaran_jump_2023}. Using the modified InfoNCE objective \cite{zhang_contrastive_2020,radford_learning_2021} and a gated graph neural network (GGNN) molecule encoder, we first show the effects of pretraining dataset sizes on morphology retrieval tasks. Transfer learning performances of GGNN molecule encoder pretrained with MoCoP is benchmarked on QSAR modeling task with varying training set sizes using the ChEMBL20 dataset \cite{gaulton_chembl_2012}. Finally, we demonstrate positive transfer of pretrained GGNNs on internal GSK pharmacokinetic data consisting of four different in vitro clearance assays.

\section{Background}

\paragraph{Learning multi-modal molecule and morphology representation with contrastive learning}
Contrastive learning is a member of the metric learning family which aims to learn an embedding space that pulls similar data together and pushes dissimilar data apart. Contrastive learning has experienced a resurgence in interest due to major advances in self-supervised learning. More recently, it has been increasingly employed to learn multi-modal data representation \cite{zhang_contrastive_2020,desai_virtex_2021,radford_learning_2021}. For MoCoP, we employ a symmetric variant of InfoNCE loss for pretraining following prior works \cite{zhang_contrastive_2020,radford_learning_2021}.

Intuitively, we aim to simultaneously learn a molecular encoder $f^{mol}$ and a morphology encoder $f^{morph}$ by minimizing the modified InfoNCE loss. Specifically, the pretraining dataset consists of $N$ molecule-morphology pairs, defined as $ \{ (\mathbf{x}^{mol}_i,  \mathbf{x}^{morph}_i )\,|\, i \in \{1,...,N\}\} $. The $i$-th molecule-morphology pair $\mathbf{x}^{mol}_i$ and $\mathbf{x}^{morph}_i$ are first encoded by their corresponding encoders $f^{mol}$ and $f^{morph}$ to produce their respective representations

\[ \mathbf{h}^{mol}_i = f^{mol}(\mathbf{x}^{mol}_i)\]
\[ \mathbf{h}^{morph}_i = f^{morph}(\mathbf{x}^{morph}_i)\].

where $\mathbf{h}^{mol}_i \in \mathbb{R}^{d^{mol}}$ and $\mathbf{h}^{morph}_i \in \mathbb{R}^{d^{morph}}$ are the encoded representations of $\mathbf{x}^{mol}_i$ and $\mathbf{x}^{morph}_i$. Each encoder representation is transformed using projection functions $g$ following

\[ \mathbf{u}^{mol}_i = g^{mol}(\mathbf{h}^{mol}_i)\]
\[ \mathbf{u}^{morph}_i = g^{morph}(\mathbf{h}^{morph}_i)\]

where $\mathbf{u}^{mol}_i \in \mathbb{R}^{proj}$ and $\mathbf{u}^{morph}_i \in \mathbb{R}^{proj}$ are vectors in the muli-modal embedding space. During training, $f^{mol}$, $f^{morph}$, $g^{mol}$, and $g^{morph}$ are jointly optimized to minimize the loss function

\[ \mathcal{L} = \alpha \cdot \mathcal{L}_{mol\rightarrow morph} + (1-\alpha) \cdot \mathcal{L}_{morph\rightarrow mol} \]

where $\alpha$ is a weighting term and $\mathcal{L}_{mol\rightarrow morph}$ and $\mathcal{L}_{morph\rightarrow mol}$ are molecule- and morphology-specific InfoNCE losses, defined as
\[ \mathcal{L}_{mol\rightarrow morph} = \frac{1}{N} \sum_{i=1}^{N} \textrm{log} \frac{e^{\langle \mathbf{u}^{mol}_i, \mathbf{u}^{morph}_i \rangle/\tau}}{\sum_{k=1}^{N} e^{\langle\mathbf{u}^{mol}_i,\mathbf{u}^{morph}_k \rangle}} \]
\[ \mathcal{L}_{morph\rightarrow mol} = \frac{1}{N} \sum_{i=1}^{N} \textrm{log} \frac{e^{\langle \mathbf{u}^{mol}_i,\mathbf{u}^{morph}_i \rangle/\tau}}{\sum_{k=1}^{N} e^{\langle\mathbf{u}^{mol}_k,\mathbf{u}^{morph}_i \rangle}} \]

with $\langle \mathbf{u},\mathbf{v} \rangle$ denoting the cosine similarity between vectors $\mathbf{u}$ and $\mathbf{v}$, and $\tau$ denotes a temperature scaling parameter.

Minimizing $\mathcal{L}$ produces encoders $f^{mol}$ and $f^{morph}$ that maximally preserve the mutual information between representations $\mathbf{h}^{mol}_i$ and $\mathbf{h}^{morph}_i$. The resulting $f^{mol}$ is then fine-tuned on downstream tasks for transfer learning.

\section{Methods}
\label{methods}
\paragraph{JUMP-CP dataset} We use a subset of the dataset \textit{cpg0016-jump}, available from the Cell Painting Gallery on the Registry of Open Data on AWS (\textit{https://registry.opendata.aws/cellpainting-gallery/}) as part of the JUMP-CP Consortium \cite{chandrasekaran_jump_2023}. This subset (as of February 2023) contains approximately 700K morphological profiles of 120K compounds in U2OS cells collected across 12 data generating centers.

Throughout our experiments, we use the precomputed well-level profiles provided with JUMP-CP. Each feature in a well-level profile is scaled independently using median and interquartile range statistics of the plate that the well belongs to. More concretely, the $i$-th feature of profile $x \in \mathbb{R}^d$ belonging to plate $p$ – denoted as $x_{i,p}$ – is preprocessed as followed 

\[x_{i,p}^{processed} = \frac{x_{i,p}^{raw} - med(X_{i,p})}{IQR(X_{i,p})} \] 

Where $x_{i,p}^{raw}$ denotes the raw feature value, $X_{i,p}$ denotes the vector of all $i$-th features in plate $p$, and $med$ and $IQR$ denote the median and interquartile range. 

We follow Way et al. \yrcite{way_predicting_2021} and remove features with low variance, features with extreme outlier values, and any blacklisted CellProfiler features that are known to be noisy unreliable \cite{way_blocklist_2019}. This results in the final set of 3,475 features.

\paragraph{ChEMBL20 dataset} We use the ChEMBL20 dataset processed by Mayr et al. \yrcite{mayr_large-scale_2018} to evaluate transfer learning. The dataset has been used extensively to evaluate and benchmark machine learning approaches for QSAR modeling \cite{wu_moleculenet_2018,yang_analyzing_2019,nguyen_meta-learning_2020}. In short, the dataset consists of approximately 450K compounds, each with sparse annotations of 1,310 binary downstream tasks spanning ADME, toxicity, physicochemical, binding, and functional. 
\paragraph{Internal GSK pharmacokinetic dataset} Internal rodent in vitro metabolism data were collated from four different intrinsic clearance assay protocols: rat liver microsomes ($CL_{int}^{RLM}$), mouse liver microsomes ($CL_{int}^{MLM}$), rat hepatocytes ($CL_{int}^{RH}$), and mouse hepatocytes ($CL_{int}^{MH}$). We convert all readouts to intrinsic clearance based on percent hepatic blood flow (PHBF) and aggregate replicate experiments for the same compound and protocol by taking the median reported PHBF. This yielded a dataset of 105,172 unique compounds with available data across all four endpoints. Finally, the data is binarized based on the median PHBF value per endpoint.

\begin{figure}[ht]
\begin{center}
\includegraphics[width=\columnwidth]{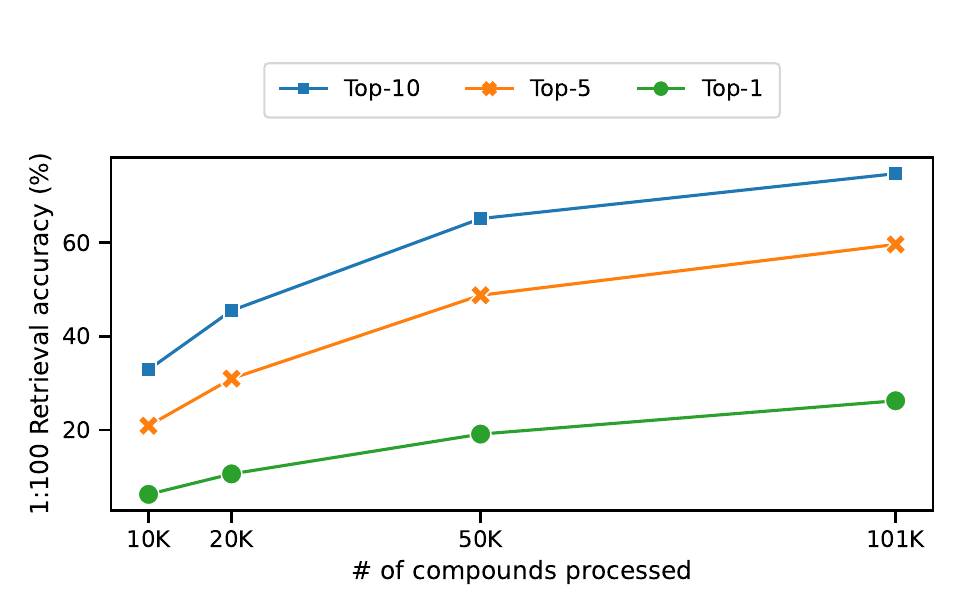}
\includegraphics[width=\columnwidth]{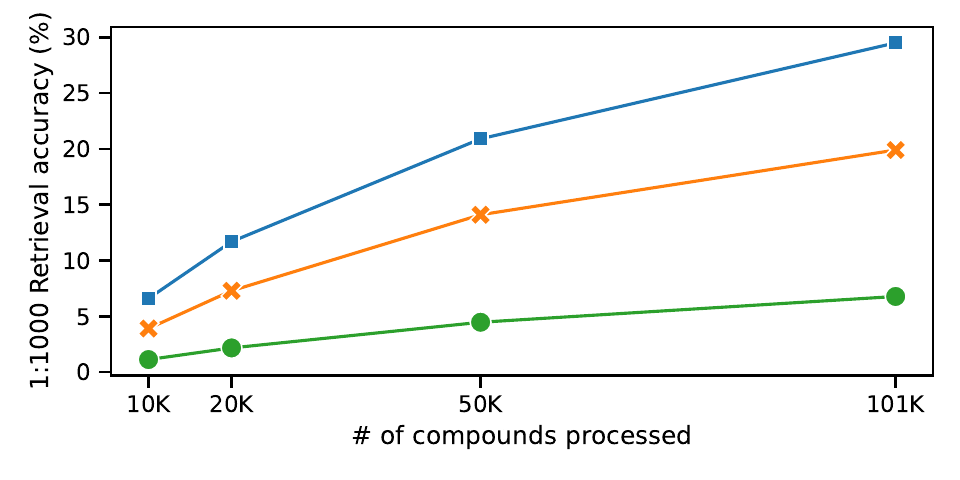}
\vskip -0.1in
\caption{Molecule and morphology retrieval performance at positive-to-negative sampling ratio of 1:100 (top) and 1:1000 (bottom) using MoCoP trained with increasing number of compounds in JUMP-CP. Average top-$k$ accuracy of retrieving molecule given morphology and vice versa is reported for $k \in \{1,5,10\}$ for each sampling ratio.}
\label{figure:retrieval}
\end{center}
\vskip -0.1in
\end{figure}
\paragraph{Contrastive pretraining procedure} Following notations from Section 2, $f_{mol}$ and $f_{morph}$ are a GGNN and a feedforward neural network (FFNN), respectively, while both $g_{mol}$ and $g_{morph}$ are single feedforward layers. Following Zhang et al. \yrcite{zhang_contrastive_2020}, $g_{mol}$ and $g_{morph}$ are non-linear transformations utilizing ReLU as the activation function. 

The model is trained for 1,000 epochs -- approximately 400,000 steps -- with a batch size of 256 on approximately 100K of the 120K compounds and 600K of the 700K morphological profiles. We follow the protocol proposed by CLIP \cite{radford_learning_2021} and OpenCLIP \cite{cherti_reproducible_2022} to use the AdamW optimizer \cite{loshchilov_decoupled_2019} with a learning rate of $10^{-3}$ and cosine annealing learning rate scheduler with 50 warm-up epochs. MoCoP hyperparameters are further detailed in Appendix \ref{appendix:mocophp}.

\paragraph{Transfer learning} We explore two transfer learning strategies for MoCoP: linear probe and fine-tuning whole model, which we refer to as MoCoP-LP and MoCoP-FT respectively. We use the Adam optimizer \cite{kingma_adam_2017} with a learning rate of $5  \times 10^{-5}$ and a batch size of 128 for both strategies.

\paragraph{Baselines} We include two baselines: training from scratch and fine-tuning from GGNNs pretrained with multitask supervised learning, which we refer to as FS and Multitask-FT, respectively.

Hyperparameter optimization is performed to ensure FS baseline is competitive. Specifically we use ChEMBL 5\% and down-sampled GSK phamacokinetic datasets to carry out a random search consisting of 50 parallel trials spanning the search space described in Appendix \ref{appendix:fshpopt} to maximize validation performance. The down-sampling procedure is detailed in Section \ref{results}.

For Multitask-FT, we first pretrain GGNNs to directly predict morphological profiles in a multi-task regression setting. Pretraining hyperparameters are optimized using random search consisting of 20 trials while fine-tuning hyperparameters are hand-tuned for performances on validation set of ChEMBL 5\%.

\paragraph{Code availability}
The source code for MoCoP is available at \href{https://github.com/GSK-AI/mocop}{https://github.com/GSK-AI/mocop}.

\begin{table*}[th]
  \caption{Performance on held-out test sets of different subsets of ChEMBL20 averaged across all tasks. FS baseline: GGNNs trained from scratch; Multitask-FT baseline: Fine-tuning GGNNs pretrained using multitask supervised learning and fine-tuned; MoCoP-LP: Linear probe on GGNNs pretrained with MoCoP; MoCoP-FT: Fine-tuning GGNNs pretrained with MoCoP. Mean and standard deviation are obtained from 9 repeats from 3 splits and 3 seeds (see Section \ref{methods} for details). The best and second best values are in bold and regular text, respectively.}
  \label{table:chembl20}
  \begin{center}
  \begin{small}
  \begin{sc}
  \centering
  \begin{tabular}{llcccc}
    \toprule
    \textbf{Metric}  & \textbf{Dataset} & \textbf{FS} & \textbf{Multitask-FT} & \textbf{MoCoP-LP} & \textbf{MoCoP-FT} \\
    \midrule\midrule
     \multirow{6}{*}{AUROC} &ChEMBL20 - 1\%      & $\color{gray}0.511\pm0.008$ & $\color{gray}0.508\pm0.007$ & $\mathbf{0.545\pm0.017}$ & $0.542\pm 0.010$  \\
      					  & ChEMBL20 - 5\%     & $\color{gray}0.571\pm0.010$ & $\color{gray}0.574\pm0.004$ & $\mathbf{0.624\pm0.018}$ & $0.621\pm 0.022$  \\
      					  & ChEMBL20 - 10\%   & $\color{gray}0.597\pm0.014$ & $\color{gray}0.588\pm0.009$ & $0.638\pm0.017$ & $\mathbf{0.646\pm 0.021}$  \\
      					  & ChEMBL20 - 25\%   & $\color{gray}0.648\pm0.017$ & $\color{gray}0.643\pm0.020$ & $0.678\pm0.015$ & $\mathbf{0.689\pm 0.018}$  \\
      					  & ChEMBL20 - 50\%   & $\color{gray}0.669\pm0.016$ & --- & --- & $\mathbf{0.693\pm 0.030}$  \\
      					  & ChEMBL20 - 100\% & $\color{gray}0.706\pm0.022$ & --- & --- &  $\mathbf{0.721\pm 0.020}$  \\
     \midrule
     \multirow{6}{*}{AUPRC} &ChEMBL20 - 1\%      & $\color{gray}0.487\pm0.013$ & $\color{gray}0.482\pm0.015$ & $\mathbf{0.511\pm0.024}$ & $0.510\pm 0.016$  \\
      					  & ChEMBL20 - 5\%     & $\color{gray}0.528\pm0.010$ & $\color{gray}0.525\pm0.013$ & $\mathbf{0.576\pm0.026}$ & $0.569\pm 0.023$  \\
      					  & ChEMBL20 - 10\%   & $\color{gray}0.550\pm0.022$ & $\color{gray}0.539\pm0.023$ & $0.588\pm0.032$ & $\mathbf{0.597\pm 0.036}$  \\
      					  & ChEMBL20 - 25\%   & $\color{gray}0.600\pm0.028$ & $\color{gray}0.595\pm0.026$ & $0.623\pm0.027$ & $\mathbf{0.640\pm 0.031}$  \\
      					  & ChEMBL20 - 50\%   & $\color{gray}0.623\pm0.026$ & --- & --- & $\mathbf{0.654\pm 0.037}$  \\
      					  & ChEMBL20 - 100\% & $\color{gray}0.662\pm0.033$ & --- & --- & $\mathbf{0.681\pm 0.033}$  \\
    \bottomrule
  \end{tabular}
  \end{sc}
\end{small}
\end{center}
\end{table*}
\section{Experimental Results and Discussion}
\label{results}
\paragraph{Scaling MoCoP to JUMP-CP}

We first evaluate if MoCoP is feasible with the JUMP-CP dataset following procedure detailed in Section \ref{methods}. Similar approaches have been previously carried out on smaller datasets collected at a single site \cite{sanchez-fernandez_contrastive_2022,zheng_cross-modal_2022}, and the aim is to test its scalability on a larger and multi-site dataset. To evaluate the pretraining performance, the accuracy of molecule and morphology retrieval is measured. Specifically, the average top-$k$ accuracy -- where $k$ can be 1, 5, or 10 -- of retrieving molecule given morphology and vice versa is reported. The positive-to-negative sampling ratio is set to 1:100 and 1:1000. 

Shown in Figure \ref{figure:retrieval}, the performance of pretraining improves as more compounds are included in the training process. The trend continues even beyond the maximum of 101K compounds, indicating pretraining can further benefit from obtaining more data. This observation highlights the importance of large public repositories of cellular imaging data. Additionally, we present training and validation curves in Appendix \ref{appendix:training}, which demonstrates a stable and convergent training process. 

Moreover, we have not extensively explored preprocessing pipelines for morphological profiles, and we anticipate that employing more advanced approaches to mitigate batch effects could improve performance.

\paragraph{Transfer learning performances on ChEMBL20}
We aim to evaluate the quality of pretrained GGNN molecule encoder by using ChEMBL20 as the downstream task. Random splits based on compounds are carried out at an 80/10/10 ratio for training, validation, and test sets. For each split, we further subsample 1\%, 5\%, 10\%, and 25\%, and 50\% of the training set to simulate an increasingly sparse data regime. 

Table \ref{table:chembl20} shows transfer learning performance on ChEMBL20. We report performance averaged across all tasks following existing works utilizing this dataset \cite{mayr_large-scale_2018,wu_moleculenet_2018,yang_analyzing_2019}. Our results indicate that fine-tuning GGNNs pretrained with MoCoP (MoCoP-FT) consistently outperformed training-from-scatch (FS) baseline across all data regimes. This improvement is also observed by simply applying a linear probe on the frozen molecule encoder (MoCoP-LP). We also observe that MoCoP-LP outperforms MoCoP-FT in lower data regime. Notably, we encounter challenges with Multitask-FT, in which GGNNs are first trained to directly predict morphological features in a multi-task regression setting. This approach fails to produce any improvements over FS baseline. Our finding is consistent with previous research that highlights the superior learning efficiency of contrastive objectives over predictive objectives.\cite{chen_uniter_2020,tian_contrastive_2020,radford_learning_2021}.

\begin{table}
  \caption{Performance on held-out test sets of GSK internal pharmacokinetic data. Mean and standard deviation are obtained from 9 repeats from 3 splits and 3 seeds (see Section \ref{methods} for details). The best values are in bold text.}
  \label{table:pkfull}
  \begin{center}
  \begin{small}
  \begin{sc}
  \centering
  \begin{tabular}{llcccc}
    \toprule
    \textbf{Metric}  & \textbf{Dataset} & \textbf{FS} & \textbf{MoCop-FT} \\
    \midrule\midrule
     \multirow{5}{*}{AUROC} & $CL_{int}^{RH}$ & $0.762\pm0.008$ & $\mathbf{0.788\pm 0.014}$  \\ %1
      					  & $CL_{int}^{MH}$   & $0.763\pm0.031$ & $\mathbf{0.791\pm 0.026}$  \\ %3
      					  & $CL_{int}^{RLM}$ & $0.845\pm0.011$ & $\mathbf{0.864\pm 0.013}$  \\ %0
      					  & $CL_{int}^{MLM}$ & $0.839\pm0.018$ & $\mathbf{0.852\pm 0.024}$  \\ %2
					  & Average & $0.802\pm0.013$ & $\mathbf{0.824\pm 0.014}$  \\ %2
     \midrule
     \multirow{4}{*}{AUPRC} & $CL_{int}^{RH}$ & $0.760\pm0.023$ & $\mathbf{0.790\pm 0.030}$  \\ %1
      					  & $CL_{int}^{MH}$   & $0.775\pm0.030$ & $\mathbf{0.795\pm 0.031}$  \\ %3
      					  & $CL_{int}^{RLM}$ & $0.851\pm0.006$ & $\mathbf{0.870\pm 0.004}$  \\ %0
      					  & $CL_{int}^{MLM}$ & $0.831\pm0.009$ & $\mathbf{0.845\pm 0.014}$  \\ %2
					  & Average & $0.804\pm0.011$ & $\mathbf{0.825\pm 0.014}$  \\ %2
    \bottomrule
  \end{tabular}
  \end{sc}
\end{small}
\end{center}
\end{table}

\paragraph{Transfer learning performances on internal GSK pharmacokinetic data}

\begin{table}
  \caption{Performance on held-out test sets of GSK internal pharmacokinetic data with down-sampled training data. Mean and standard deviation are obtained from 9 repeats from 3 splits and 3 seeds (see Section \ref{methods} for details). The best values are in bold text.}
  \label{table:pkds}
  \begin{center}
  \begin{small}
  \begin{sc}
  \centering
  \begin{tabular}{llcccc}
    \toprule
    \textbf{Metric}  & \textbf{Dataset} & \textbf{FS} & \textbf{MoCop-FT} \\
    \midrule\midrule
     \multirow{5}{*}{AUROC} & $CL_{int}^{RH}$ & $0.716\pm0.046$ & $\mathbf{0.763\pm 0.057}$  \\ %1
      					  & $CL_{int}^{MH}$   & $0.716\pm0.056$ & $\mathbf{0.805\pm 0.049}$  \\ %3
      					  & $CL_{int}^{RLM}$ & $0.800\pm0.011$ & $\mathbf{0.824\pm 0.018}$  \\ %0
      					  & $CL_{int}^{MLM}$ & $0.779\pm0.015$ & $\mathbf{0.805\pm 0.023}$  \\ %2
					  & Average & $0.752\pm0.028$ & $\mathbf{0.799\pm 0.033}$  \\ %2
     \midrule
     \multirow{4}{*}{AUPRC} & $CL_{int}^{RH}$ & $0.715\pm0.053$ & $\mathbf{0.768\pm 0.049}$  \\ %1
      					  & $CL_{int}^{MH}$   & $0.710\pm0.044$ & $\mathbf{0.799\pm 0.046}$  \\ %3
      					  & $CL_{int}^{RLM}$ & $0.820\pm0.011$ & $\mathbf{0.842\pm 0.018}$  \\ %0
      					  & $CL_{int}^{MLM}$ & $0.818\pm0.019$ & $\mathbf{0.846\pm 0.027}$  \\ %2
					  & Average & $0.766\pm0.025$ & $\mathbf{0.814\pm 0.031}$  \\ %2
    \bottomrule
  \end{tabular}
  \end{sc}
\end{small}
\end{center}
\end{table}
The quality of pretrained GGNNs is further evaluated using a subset of GSK internal pharmacokinetic data as downstream tasks. This dataset consists of 4 tasks as detailed in Section \ref{methods}. Unlike the previous experiment with ChEMBL20, here we employ scaffold splitting, which has been shown to provide better estimates of model performances in QSAR tasks \cite{kearnes_modeling_2017,wu_moleculenet_2018}. The compounds are first clustered using the Butina algorithm implemented in RDKit with a Euclidean distance function and a distance cutoff of 0.6. The clusters are ordered by size, and for every of six clusters, four are assigned to the training set, one to the validation set, and one to the test set. The procedure is repeated with random cluster ordering to create two additional splits. For each split, a down-sampled version is created randomly selecting a single compound from each cluster to uniformly sample the chemical space in our dataset. 

Using results from the previous experiment, we benchmark the most performant approach MoCoP-FT, where each model is repeated 9 times with 3 splits and 3 seeds. We again observe that MoCoP-FT consistently outperforms FS baseline across both full and down-sampled datasets, shown in Table \ref{table:pkfull} and \ref{table:pkds}, respectively. On the full dataset, pretrained GGNNs show an average improvement of 2.6\% in AUPRC across the 4 individual tasks. This effect is increased to 6.3\% in AUPRC when less data is available for training. We expect performance can be further improved by considering using related endpoints as descriptors, as demonstrated by Broccatelli et al. \yrcite{broccatelli_benchmarking_2022}.

This result offers a glimpse at the potential of using datasets not directly related to the learning task at hand in improving QSAR models. While the results in this study are limited to a single publicly available high-content imaging dataset, other high-dimensional readouts such as transcriptomics and proteomics can be used to augment QSAR modeling in similar manners. Further investigation of routine re-use of high-dimensional data in standard QSAR workflows is warranted in future works.

\section{Conclusion}

In this study, we explore MoCoP as a means to improve the performance of QSAR models. We scale MoCoP to approximately 100K molecules and 600K morphological profiles, and evaluate pretrained GGNNs molecule encoder on both public and internal downstream tasks.

Our results demonstrate that MoCoP consistently improves the performance of GGNNs in QSAR tasks, especially in low-data regimes when compared to training-from-scratch and multitask supervised pretraining baselines. We observe this trend in both the ChEMBL20 dataset and GSK internal pharmacokinetic data, indicating that the approach is applicable across a range of datasets and tasks.

Our work also suggests that data from unbiased high-dimensional assays, beyond cellular imaging, can improve QSAR models via contrastive pretraining. Future works will further explore this approach with other data sources such as transcriptomics and proteomics. Overall, we believe our work can be combined with existing methods to improve model performances and expands the deep learning toolbox available for QSAR applications.

\bibliography{mocop}
\bibliographystyle{icml2023}

%%%%%%%%%%%%%%%%%%%%%%%%%%%%%%%%%%%%%%%%%%%%%%%%%%%%%%%%%%%%%%%%%%%%%%%%%%%%%%%
%%%%%%%%%%%%%%%%%%%%%%%%%%%%%%%%%%%%%%%%%%%%%%%%%%%%%%%%%%%%%%%%%%%%%%%%%%%%%%%
% APPENDIX
%%%%%%%%%%%%%%%%%%%%%%%%%%%%%%%%%%%%%%%%%%%%%%%%%%%%%%%%%%%%%%%%%%%%%%%%%%%%%%%
%%%%%%%%%%%%%%%%%%%%%%%%%%%%%%%%%%%%%%%%%%%%%%%%%%%%%%%%%%%%%%%%%%%%%%%%%%%%%%%

\newpage
\appendix
\onecolumn
\setcounter{figure}{0}   
\setcounter{table}{0}   
\section{FS Baseline Hyperparameter Tuning}
\label{appendix:fshpopt}
Hyperparameter optimization is done on the search space below using a random search consisting of 50 parallel trials. Bold and underscored values denote the selected hyperparameters for ChEMBL20 and pharmacokinetic data, respectively.
\begin{table}[h]
\begin{center}
\begin{small}
\begin{sc}
\centering
\begin{tabular}{lc}
\toprule
\textbf{Hyperparameter}  & \textbf{Search space} \\
\midrule\midrule
Learning rate & \{$10^{-5}$, $5 \times 10^{-5}$,$10^{-4}$, $5 \times 10^{-4}$, \underline{$\mathbf{10^{-3}}$} \} \\
\# of GGNN layers & {4, 5, 6, \underline{\textbf{7}}, 8} \\
Batch size & {64, \textbf{128}, \underline{256}, 512} \\
\bottomrule
\end{tabular}
\end{sc}
\end{small}
\end{center}
\end{table}
\section{Training MoCoP}
\subsection{Hyperparameters}
\label{appendix:mocophp}
MoCoP hyperparameters used in this work are provided in table \ref{table:mocophp} below.
\begin{table}[h]
\label{table:mocophp}
\begin{center}
\begin{small}
\begin{sc}
\centering
\begin{tabular}{lc}
\toprule
\multicolumn{2}{c}{\textbf{Hyperparameter}} \\ 
\midrule\midrule
\# of GGNN layers in $f^{mol}$ & 6 \\
FF layers dimensions in $f^{morph}$ & {[512, 256, 128]} \\
$d^{mol}$ & 1024 \\
$d^{morph}$ & 128 \\
$d^{proj}$ & 128 \\
Learning rate & $10^{-3}$ \\
Learning rate scheduler & Cosine annealing with linear warm-up \\
\# of warm-up epochs & 50 \\
\# of epochs & 1,000 \\
Batch size & 256 \\
\bottomrule
\end{tabular}
\end{sc}
\end{small}
\end{center}
\end{table}

\subsection{Training}
\label{appendix:training}
We develop a simple sampling procedure to accommodate the one-to-many nature of molecule-to-morphology mapping. Specifically, for each batch of size $N$, we first randomly select $N$ unique compounds, and for each compound randomly select a single corresponding morphology. We detail the procedure in Algorithm \ref{alg:sampling}.

\begin{algorithm}[h]
   \caption{MoCoP Batch Sampling}
   \label{alg:sampling}
\begin{algorithmic}
   \STATE {\bfseries Input:}
   \STATE \hskip1em Batch size $N$
   \STATE \hskip1em Number of unique molecules $K$
   \STATE \hskip1em All unique molecules $\mathbf{X}^{mol} = \{\mathbf{x}^{mol}_i\,|\, i \in \{1, ..., K\}\}$
   \STATE \hskip1em Mapping of unique molecules to corresponding morphologies $M=\{(\mathbf{x}^{mol}_i, \mathbf{X}^{morph}_i)\,|\, i \in \{1, ..., K\}\}$
   \STATE
   \STATE $batch \gets \{\}$
   \FOR{$i=1$ {\bfseries to} $N$}
   \STATE Sample $\mathbf{x}^{mol}_i$ from $\mathbf{X}^{mol}$
   \STATE Collect corresponding $\mathbf{X}^{morph}_i$ from mapping $M$
   \STATE Sample $\mathbf{x}^{morph}_i$ from $\mathbf{X}^{morph}_i$
   \STATE $\mathbf{X}^{mol} \gets \mathbf{X}^{mol} \setminus \{\mathbf{x}^{mol}\}$
   \STATE $batch \gets batch \cup \{(\mathbf{x}^{mol}_i, \mathbf{x}^{morph}_i)\}$
   \ENDFOR
   \STATE {\bfseries Return}  $batch$
\end{algorithmic}
\end{algorithm}

\begin{figure}[H]
\begin{center}
\includegraphics[width=0.5\textwidth]{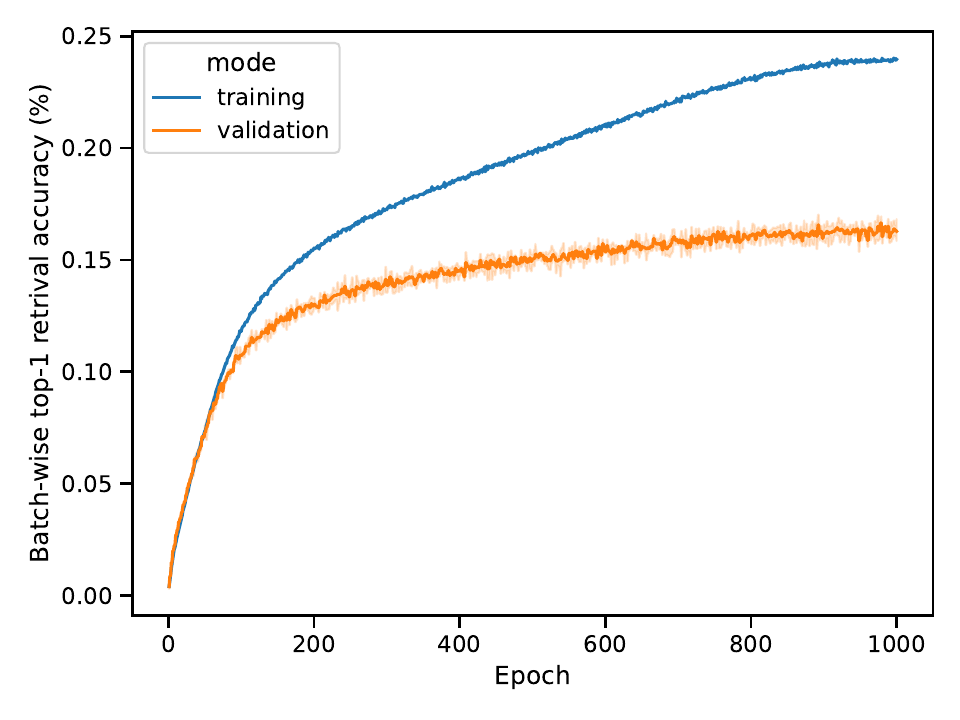}
\vskip -0.1in
\caption{Training and validation curves of MoCoP across 3 different JUMP-CP splits and random initializations. The reported metric is calculated as the average top-1 accuracy for retrieving molecule and morphology in a batch.}
\label{figure:train}
\end{center}
\vskip -0.1in
\end{figure}

The sampling procedure above produces stable training behaviors for MoCoP, demonstrated in the training and validation curves in Figure \ref{figure:train}. Training is carried out on a single NVIDIA V100 GPU over 6 days.

\section{Effects of Batch Size and $d^{proj}$ on Transfer Learning}
A small hyperparameters study was conducted to investigate the effects of batch size and $d^{proj}$ on transfer learning performance using the down-sampled GSK pharmacokinetic dataset as the downstream tasks. We observe that smaller batch size produces more transferable molecule encoder while $d^{proj}$ does not significantly affect overall performances.
\begin{table}[h]
  \label{table:bs}
  \begin{center}
  \begin{small}
  \begin{sc}
  \centering
  \begin{tabular}{llcccc}
    \toprule
    \multirow{2}{*}{\textbf{Metric}}  & \multirow{2}{*}{\textbf{Dataset}} & \textbf{Batch size = 1024} & \textbf{Batch size = 512} & \textbf{Batch size = 256} \\
    						  &						    & \textbf{$d^{proj}$ = 128}   & \textbf{$d^{proj}$ = 128}	& \textbf{$d^{proj}$ = 128} \\
    \midrule\midrule
     \multirow{5}{*}{AUROC} & $CL_{int}^{RH}$ & $0.747\pm 0.077$ & $0.756\pm0.053$ & $\mathbf{0.780\pm 0.080}$  \\ %1
      					  & $CL_{int}^{MH}$   & $0.772	\pm0.045$ & $0.801\pm0.042$ & $\mathbf{0.831\pm 0.066}$  \\ %3
      					  & $CL_{int}^{RLM}$ & $0.817\pm0.008$ & $0.825\pm0.008$ & $\mathbf{0.836\pm 0.030}$  \\ %0
      					  & $CL_{int}^{MLM}$ & $0.791\pm0.013$ & $0.796\pm0.003$ & $\mathbf{0.816\pm 0.039}$  \\ %2
					  & Average & $0.782\pm 0.032$ & $0.795\pm 0.024$ & $\mathbf{0.815\pm 0.053}$  \\ %2
     \midrule
     \multirow{4}{*}{AUPRC} & $CL_{int}^{RH}$ & $0.741\pm 0.059$ & $0.764\pm 0.048$ & $\mathbf{0.779\pm 0.071}$  \\ %1
      					  & $CL_{int}^{MH}$   & $0.761\pm 0.013$ & $0.794\pm 0.010$ & $\mathbf{0.841\pm 0.062}$  \\ %3
      					  & $CL_{int}^{RLM}$ & $0.831\pm 0.002$ & $0.840\pm 0.013$ & $\mathbf{0.856\pm 0.027}$  \\ %0
      					  & $CL_{int}^{MLM}$ & $0.829\pm 0.024$ & $0.834\pm 0.017$ & $\mathbf{0.857\pm 0.034}$  \\ %2
					  & Average & $0.800\pm 0.023$ & $0.808\pm 0.016$ & $\mathbf{0.833\pm 0.046}$  \\ %2
    \bottomrule
  \end{tabular}
  \end{sc}
\end{small}
\end{center}
\end{table}

\begin{table}[h]
%  \caption{Performance on held-out test sets of GSK internal pharmacokinetic data. Mean and standard deviation are obtained from 9 repeats from 3 splits and 3 seeds (see Section \ref{methods} for details). The best values are in bold text.}
  \label{table:dproj}
  \begin{center}
  \begin{small}
  \begin{sc}
  \centering
  \begin{tabular}{llcccc}
    \toprule
    \multirow{2}{*}{\textbf{Metric}}  & \multirow{2}{*}{\textbf{Dataset}} & \textbf{$d^{proj}$ = 128}    & \textbf{$d^{proj}$ = 256}   & \textbf{$d^{proj}$ = 512} \\
    						  &						    & \textbf{Batch size = 1024} & \textbf{Batch size = 1024} & \textbf{Batch size = 1024}\\
    \midrule\midrule
     \multirow{5}{*}{AUROC} & $CL_{int}^{RH}$ & $0.747\pm 0.077$ & $\mathbf{0.745\pm 0.068}$ & $\mathbf{0.737\pm 0.057}$  \\ %1
      					  & $CL_{int}^{MH}$   & $0.772	\pm0.045$ & $\mathbf{0.819\pm 0.049}$ & $\mathbf{0.819\pm 0.004}$  \\ %3
      					  & $CL_{int}^{RLM}$ & $0.817\pm0.008$ & $0.817\pm 0.002$ & $\mathbf{0.819\pm 0.003}$  \\ %0
      					  & $CL_{int}^{MLM}$ & $0.791\pm0.013$ & $\mathbf{0.806\pm 0.009}$ & $0.803\pm 0.017$  \\ %2
					  & Average & $0.782\pm 0.032$ & $0.794\pm 0.031$ & $\mathbf{0.815\pm 0.053}$  \\ %2
     \midrule
     \multirow{4}{*}{AUPRC} & $CL_{int}^{RH}$ & $0.741\pm 0.059$ & $\mathbf{0.758\pm 0.058}$ & $0.735\pm 0.066$  \\ %1
      					  & $CL_{int}^{MH}$   & $0.761\pm 0.013$ & $0.774\pm 0.029$ & $\mathbf{0.803\pm 0.043}$  \\ %3
      					  & $CL_{int}^{RLM}$ & $0.831\pm 0.002$ & $0.833\pm 0.004$ & $\mathbf{0.837\pm 0.010}$  \\ %0
      					  & $CL_{int}^{MLM}$ & $0.829\pm 0.024$ & $\mathbf{0.843\pm 0.015}$ & $0.840\pm 0.011$  \\ %2
					  & Average & $0.800\pm 0.023$ & $0.802\pm 0.025$ & $\mathbf{0.803\pm 0.010}$  \\ %2
    \bottomrule
  \end{tabular}
  \end{sc}
\end{small}
\end{center}
\end{table}

%
%\begin{itemize}
%\item[--] Number of GGNN layers: $\{3, 7, 9\}$
%\item[--] Fully connected layer dimension: $\{1024, 2048\}$
%\item[--] Batch size: $\{128, 256, 512\}$
%\item[--] Learning rate: ${10^{\{-4.0,-3.75, -3.5, -3.25\}}}$
%\end{itemize}

\end{document}